\documentclass[english]{article}

\usepackage{color}
\makeatletter
\usepackage[latin1]{inputenc}
\usepackage{graphicx}

\usepackage{geometry}
\usepackage{color}

\usepackage{amstext}
\usepackage{amssymb}
\usepackage{color}

\makeatother

\usepackage{babel}

\begin{document}

\title{A field-theoretical approach to non-relativistic QED.}
\author{Ladislaus Alexander B\'anyai*  and Mircea Bundaru**}

\maketitle

{\sl Institut f\"ur Theoretische Physik, Goethe-Universit\"at, Frankfurt am Main*

National Institute of Physics and Nuclear Engineering - Horia Hulubei, Bucharest** }

\begin{abstract}
We give here a field-theoretical derivation  of the Hamiltonian of the non-relativistic quantum electrodynamics in the Coulomb gauge using the Lagrange formalism. It leads to the same result as the usual derivation, where one just replaces the classical vector potential in the minimal coupling of the second quantized electron Hamiltonian by the quantized one and adds the photon energy. 
This approach illustrates however  the proper use of the Euler-Lagrange equations and the canonical formalism that fail  if one tries to build up the many-body quantum theory starting from the classical theory of point-like particles interacting with  electromagnetic fields.
\end{abstract}

\maketitle
\section{Introduction}
The traditional textbook description of the quantum mechanical interaction of charged particles starts directly with a Hamiltonian of Coulomb interacting  particles omitting the divergent self-interaction. This Hamiltonian is not at all self-evident. It stems not from  a classical Hamiltonian obtained from a given Lagrangian and a quantization according to the "Poisson brackets to commutators" recipe,  as it was the case for a single particle in a potential. A consistent Lagrangian formalism for classical charged point-like particles interacting with the  electromagnetic field, whose sources are these  themselves, is missing. One has to exclude by hand the self-interaction from the Lorentz force in order to obtain  meaningful equations. Nevertheless, it was a useful construction, which in its second quantized version may be justified as neglecting any correction in $1/c$.       

One has tried since almost 100 years \cite{Darwin} (see also \cite{Landau}) to construct also a version of the many-body Hamiltonian of charged particles that includes besides the Coulomb interaction also  non-relativistic electromagnetic interactions,  at least to order $1/c^2$. The chosen way  started again from the electromagnetic interaction of point-like classical charges, followed by a quantization. Lacking a consistent Lagrangian and canonical  formalism   one arrived at the misleading Darwin-Hamiltonian \cite{Darwin} of order $1/c^2$, still surviving in the literature. Surprisingly, one ignored the length existing  correct formulation  of the non-relativistic QED.  In a recent paper \cite{Banyai} it was shown, that the $1/c^2$ approximation of the later restricted to the states without photons gives rise indeed to a quantum mechanical  Hamiltonian of charged particles up to order $1/c^2$ , that could have been obtained also by the more primitive construction, provided one had chosen the physical Coulomb gauge without spurious degrees of freedom.

 In order to bring more clarity in this matter for those not familiar with the quantum field theory,  we give here a rigorous field-theoretical Lagrangian derivation of the non-relativistic QED. This is the proper quantum-mechanical  frame for many-body theories of charged particles and photons. One should avoid the vicious electromagnetism of classical  point like charges. 
    
We use here an analogous path to that used  in the relativistic quantum field theory,  describing here only the interaction between non-relativistic electrons and photons. We proceed by three steps. The first is to build up a classical Lagrangian out of classical fields including the elctromagnetic fields and  an one-electron wave function  that gives rise to the Maxwell equations coupled to the charge density and current defined by the one-electron Schr\"odinger equation. The second step is to choose the Coulomb gauge  in order to eliminate the spurious degrees of freedom that allows  to build  a classical Hamiltonian. The last step is to quantize all the physical fields (electron wave function and transverse e.m. potential) in this Hamiltonian, according to the standard quantization  recipe.  Of course, for sake of stability the many body system should contain also particles of opposite charge, but the extension of the formalism to include these is trivial.

\section{Field theory.}
In the classical field-theory one defines the action ${\cal A}$ 
\[
{\cal A}=\int d {\vec x} \int dt {\cal L}({\vec x},t)
\]
by a Lagrange density ${\cal L}({\vec x},t)$ depending on some fields $\phi_i({\vec x},t)$  and their first time and space derivatives. The variational principle $\delta A=0$ gives rise to the generalized Euler-Lagrange equations 

\[
\frac{\partial}{\partial t}\frac{\delta {\cal L}}{\delta {\dot \phi_i({\vec x}, t)}}+
\frac{\partial}{\partial x_\mu} \frac{\delta {\cal L}}{\delta 
\frac{\partial \phi_i({\vec x}, t)}{\partial x_\mu} }
-\frac{\delta {\cal L}}{\delta \phi_i({\vec x},t)}=0 \enspace .
\]
Here the symbol $\partial$ means ordinary derivation, while the symbol $\delta$ means functional derivation.  Two Lagrangian densities that differ by the time derivative  or by the divergence of a function give rise to the same action and therefore are considered to be equivalent, taking into account the vanishing of the fields at infinity.

The generalized canonical conjugate momenta for the fields $\phi_i ({\vec x},t)$ are defined by 
\[
\Pi_{\phi_i} =\frac{\delta{\cal L}}{\delta {\dot {\phi}_i}}
\]
and the Hamiltonian density  is
\[
{\cal H}(\phi,\Pi_\phi)=-{\cal L}+\Pi_{\phi_i} {\dot \phi_i} \enspace ,
\]
provided no relations (constraints) appear between the canonical conjugate momenta. Lagrangians with constraints however have to be handled with Dirac's canonical formalism \cite{Dirac1, Dirac2}, that implies also a redefinition of the Poisson bracket. 

\section{Classical Maxwell fields coupled to a quantum mechanical electron.}

The  classical Maxwell equations  are two with sources
\begin{eqnarray}
\nabla\times\vec{B} & = & \frac{4\pi}{c}\vec{j}+\frac{1}{c}\frac{\partial}{\partial t}\vec{E} \label{Max1a}
\\
\nabla\vec{E} & = & 4\pi\rho \label{Max2a}
\end{eqnarray}
and two  without sources 
\begin{eqnarray}
\nabla\vec{B} & = & 0\label{Max3a}
\\
\nabla\times\vec{E} & = & -\frac{1}{c}\frac{\partial}{\partial t}\vec{B} \enspace 
\label{Max4a}
\end{eqnarray}
(see \cite{Landau} for the chosen units).
The equations without sources are automatically satisfied by the introduction of the  electromagnetic potentials
 
\begin{eqnarray}
\vec{B} & = & \nabla\times\vec{A} \label{e.m.pot1}
\\
\vec{E} & = & -\nabla V-\frac{1}{c}\frac{\partial}{\partial t}\vec{A} \enspace .
\label{e.m.pot2}
\end{eqnarray}

Let us now suppose, that the sources
\begin{eqnarray}
\rho({\vec x},t)&=&e\psi({\vec x}, t)^* \psi({\vec x},t)
\\
\vec{j}({\vec x},t)&=&\frac{e}{2m}\psi({\vec x},t)^*\left(-\imath \hbar\nabla
+\frac{e}{ c}{\vec A}({\vec x},t) \right)\psi({\vec x},t) +c.c 
\end{eqnarray}
are given by a single electron wave function $\psi({\vec x},t)$
(for simplicity without spin),  described by the quantum mechanical Schr\"odinger equation

\begin{equation}
\imath\hbar\frac{\partial}{\partial t}\psi=\left(\frac{1}{2m}\left(-\imath\hbar\nabla+\frac{e}{c}\vec{A}(x,t)\right)^{2}+eV(x,t)\right)\psi \enspace . \label{Schro}
\end{equation}

Using only the Schr\"odinger equation the sources satisfy  the continuity equation (required also by consistency)
\begin{equation}
\nabla\vec{j}+\frac{\partial}{\partial t}\rho=0 \enspace .\label{continuity}
\end{equation}
The electromagnetic potentials and the wave function are not uniquely defined, they allow a simultaneous gauge  transformation 
\begin{eqnarray}
V({\vec x},t)&\to& V({\vec x},t)+\frac{1}{c}{\dot \chi({\vec x},t)} \label{gauge}
\\
\vec{A}({\vec x},t)&\to& \vec{A}({\vec x},t) -\nabla \chi({\vec x},t)\nonumber
\\
\psi({\vec x},t) &\to& \psi({\vec x},t)e^{-\frac{\imath e}{\hbar c}\chi({\vec x},t)}\nonumber
\end{eqnarray}
with an arbitrary continuous  function $\chi({\vec x},t)$ that do not change neither the Maxwell fields ${\vec B}$, ${\vec E}$, their sources $\rho$, ${\vec j}$,  nor the whole system of equations from Eq.\ref{Max1a} to  Eq. \ref{continuity}.

The source-less Maxwell equations Eq.\ref{Max3a} and Eq.\ref{Max4a} are satisfied automatically in terms of the electromagnetic potentials through  Eq.\ref{e.m.pot1} and Eq.\ref{e.m.pot2}. Therefore, one should concentrate only on the  Eq.\ref{Max1a}, Eq.\ref{Max2a} and Eq.\ref{Schro}. 

The first problem is then to find a Lagrangian giving rise to these equation in terms of the fields $V({\vec x},t)$, ${\vec A}({\vec x},t)$ and $\psi({\vec x},t)$ as dynamical variables (generalized coordinates).  Thereafter, one has to find the classical Hamiltonian  and at the end quantize simultaneously the wave function of the electron $\psi$ and the transverse vector potential ${\vec A}_\bot$. 

\section{Classical Lagrange density for the Maxwell equations coupled to a quantum mechanical electron.}

In our case the fields are $V({\vec x},t)$, ${\vec A}({\vec x},t)$ and $\psi({\vec x},t)$. 
It is easy to see, that the "photon" Lagrange density 
\begin{equation}
{\cal L}_{ph} ({\vec x},t)=\frac{1}{8\pi}\left(\nabla V({\vec x},t)+ \frac{1}{c} {\dot{\vec A}}({\vec x} ,t)\right)^2 -\frac{1}{8\pi}\left(\nabla \times {\vec A}({\vec x}, t)\right)^2
\label{Lph}
\end{equation}
 through the generalized  Euler-Lagrange equations for the fields  ${\vec A}({\vec x},t)$ and $V({\vec x},t)$ gives rise to the Maxwell equations Eq.\ref{Max1a} respectively
  Eq.\ref{Max2a}, however without sources.

On its turn, the Lagrangian density of the "electron"  
\begin{equation}
{\cal L}_e({\vec x},t)=-\frac{\hbar^2}{2m}\nabla \psi^*({\vec x},t)\nabla \psi({\vec x},t) -
\frac{\imath\hbar}{2}\left({\dot \psi}({\vec x},t)^*\psi({\vec x},t)-\psi({\vec x},t)^*{\dot{\psi}}({\vec x},t)\right)\label{Le}
\end{equation}
 for the field  $\psi({\vec x},t)^*$ gives rise to the free Schr\"odinger equation (uncoupled to the e.m. fields)
\begin{equation}
\imath \hbar \frac{\partial}{\partial t} \psi({\vec x},t)=-\frac{\hbar^2}{2m}\nabla^2\psi({\vec x}, t)\enspace . \label{Schr0}
\end{equation}

We shall  introduce the e.m. interaction between the electron and the electromagnetic field by the so called  "minimal way"  ensuring gauge invariance. It requires  the replacements
\begin{eqnarray}
\frac{\hbar}{\imath}\nabla\psi & \to & \left(\frac{\hbar}{\imath}\nabla +\frac{e}{c}{\vec A}\right)\psi
\\
\frac{\hbar}{\imath}\frac{\partial}{\partial t} \psi &\to &  
\left( \frac{\hbar}{\imath}\frac{\partial}{\partial t} +e V\right))\psi
\end{eqnarray}
in the electron Lagrangian. 

Thus, the total Lagrangian is 

\begin{eqnarray}
{\cal L}&=&\frac{1}{8\pi}\left(\nabla V+ \frac{1}{c}\frac{\partial}{\partial t} {\vec A}\right)^2 -\frac{1}{8\pi}\left(\nabla \times {\vec A}\right)^2 \label{Lagran}
\\
&-&\frac{1}{2m}\left(-\frac{\hbar}{\imath}\nabla +\frac{e}{c}{\vec A}\right)\psi^*
\left(\frac{\hbar}{\imath}\nabla +\frac{e}{c}{\vec A}\right)\psi \nonumber
\\
&-&\frac{1}{2}\psi^*\left(\frac{\hbar}{\imath}\frac{\partial}{\partial t} +eV\right)\psi-
\frac{1}{2}\psi\left(-\frac{\hbar}{\imath}\frac{\partial}{\partial t} + eV\right)\psi^* \nonumber \enspace .
\end{eqnarray}
This is obviously gauge invariant and gives rise to the correct coupled  equations.

\section{The classical Hamiltonian in the Coulomb gauge.}

Unfortunately, the above introduced Lagrangian density Eq.\ref{Lagran} is a so called singular one. The time derivative of the variable $V$ is not present in it and therefore the corresponding canonical momentum is vanishing i.e. we have a constraint in the canonical formalism. Lagrangians with constraints, as we already mentioned, have to be handled with Dirac's canonical formalism \cite{Dirac1, Dirac2}, that implies also a redefinition of the Poisson bracket.  
The simplest way out is however to use the choice of the gauge in such a way as to eliminate the spurious degrees of freedom from the Lagrangian before we could construct a Hamiltonian. 

The Coulomb gauge defined by 
\begin{equation}
\nabla {\vec A}({\vec x},t)=0
\end{equation} 
leaves only the physical transverse degrees of freedom of the photons and simultaneously  eliminates the scalar potential in favor of the charge density

\begin{equation}
V({\vec x},t)=\int d {\vec x}p \frac{\rho ({\vec x}p,t) }{|{\vec x}-{\vec x}p|} \enspace . \label{Coul}
\end{equation}

Taking into account, that in the Coulomb gauge we have only two real degrees of freedom for the e.m field  (the transverse vector potential  ${\vec A}_{\bot} $) and a complex degree of freedom $\psi$ for the electrons , we may define their canonical conjugate momenta as

\begin{eqnarray}
\Pi_{\psi}&\equiv & \frac{\delta  {\cal L}}{\delta {\dot \psi}} =-\frac{\imath \hbar}{2} \psi^*
\\
\Pi_{\psi^*} &\equiv& \frac{\delta { \cal L}}{\delta {\dot \psi^*}} =\frac{\imath \hbar}{2} \psi
\\
\Pi^{\mu}_{{A_\bot}}&\equiv & 
\frac{{\cal L}}{\delta {\dot A}_\bot^\mu }=
\frac{1}{4\pi c}\left(\frac{\partial}{x_\mu} V +\frac{1}{c} {\dot{ A}}_\bot^\mu\right), \quad (\mu=1^,2,3) \enspace .
\end{eqnarray}
and the Hamiltonian density  is

\begin{eqnarray}
{\cal H}& =& -{\cal L}+ {\vec \Pi_{A_{\bot}}{\dot{\vec  A}}_\bot}+ \Pi_\psi \psi +
\Pi_{\psi^*} \psi^*
\\
 &=&-\frac{1}{8\pi}\left(\nabla V+ \frac{1}{c} {\dot{\vec A}_\bot}\right)^2 +\frac{1}{8\pi}\left(\nabla \times {\vec A}_\bot \right)^2
+\frac{1}{4\pi c}{\dot {\vec A}_\bot} \left(\nabla V+ \frac{1}{c} {\dot{\vec A}_\bot}\right)
\nonumber
\\
&+&\frac{1}{2m}\left(-\frac{\hbar}{\imath}\nabla \psi^* +\frac{e}{c}{\vec A}_\bot\psi^*\right)
\left(\frac{\hbar}{\imath}\nabla\psi +\frac{e}{c}{\vec A}_\bot\psi\right) 
+eV\psi^*\psi  \enspace .\nonumber
\end{eqnarray}
or

\begin{eqnarray}
{\cal H}& =&
-\frac{1}{4\pi}\left(\nabla V+ \frac{1}{c} {\dot{\vec A}_\bot}\right)^2 +\frac{1}{8\pi}\left(\nabla \times {\vec A}_\bot \right)^2
-\frac{1}{4\pi }\nabla V \left(\nabla V+ \frac{1}{c} {\dot{\vec A}_\bot}\right)
\nonumber
\\
&+&\frac{1}{2m}\left(-\frac{\hbar}{\imath}\nabla \psi^* +\frac{e}{c}{\vec A}_\bot\psi^*\right)
\left(\frac{\hbar}{\imath}\nabla\psi +\frac{e}{c}{\vec A}_\bot\psi\right) 
+eV\psi^*\psi  \enspace .\nonumber
\end{eqnarray}
In   the  Hamiltonian
\[
H=\int d {\vec x} {\cal H}({\vec x})
\]
 one may use a partial integration in order to obtain

\begin{eqnarray*}
H &=&\int d{\vec x}\left[\frac{1}{4\pi}\left(\nabla V+ \frac{1}{c} {\dot{\vec A}_\bot}\right)^2 +\frac{1}{8\pi}\left(\nabla \times {\vec A}_\bot \right)^2
+\frac{1}{4\pi} V\nabla \left(\nabla V+ \frac{1}{c} {\dot{\vec A}_\bot}\right)\right.
\\
&+&\left.\frac{1}{2m}\left(-\frac{\hbar}{\imath}\nabla \psi^* +\frac{e}{c}{\vec A}_\bot\psi^*\right)
\left(\frac{\hbar}{\imath}\nabla\psi +\frac{e}{c}{\vec A}_\bot\psi\right) 
+eV\psi^*\psi \right] \enspace .
\end{eqnarray*}

Due to the transversality of the vector potential and expressing  the scalar  potential through the charge density Eq. \ref{Coul} one gets

\begin{eqnarray*}
H &=&\int d{\vec x}\left[\frac{1}{4\pi}\left(\nabla V+ \frac{1}{c} {\dot{\vec A}_\bot}\right)^2 +\frac{1}{8\pi}\left(\nabla \times {\vec A}_\bot \right)^2 \right.
\\
&+&\left.\frac{1}{2m}\left(-\frac{\hbar}{\imath}\nabla \psi^* +\frac{e}{c}{\vec A}_\bot\psi^*\right)
\left(\frac{\hbar}{\imath}\nabla\psi +\frac{e}{c}{\vec A}_\bot\psi\right) 
+\frac{1}{2}eV\psi^*\psi\right]  \enspace .
\end{eqnarray*}
or in terms of the canonical variables
\begin{eqnarray}
H &=&\int d{\vec x}\left[ 4\pi c^2 {\vec \Pi}_{{A_\bot}}^2
 +\frac{1}{8\pi}\left(\nabla \times {\vec A}_\bot \right)^2 \right.\label{Hamiltoniana-Coul}
\\
&+&\left.\frac{1}{2m}\left(-\frac{\hbar}{\imath}\nabla \psi^* +\frac{e}{c}{\vec A}_\bot\psi^*\right)
\left(\frac{\hbar}{\imath}\nabla\psi +\frac{e}{c}{\vec A}_\bot\psi\right) 
+\frac{1}{2}eV\psi^*\psi \right]  \nonumber
\end{eqnarray}

The first two terms are typical oscillator terms with propagation velocity $c$ and these may be brought to a diagonal form as a sum of oscillator Hamiltonians.

\section{Quantization.}
 Starting from our classical Hamiltonian in Coulomb gauge Eq.\ref{Hamiltoniana-Coul},
after the usual equal-time quantization of  the anti-commuting electron wave functions
\[
[\psi({\vec x},t),\psi^+({\vec x}p,t)]_+=\delta({\vec x}-{\vec x}p)
\]
 and the introduction of  creation and annihilation operators
  $b_{\vec{q},\lambda}^{+}$ and $b_{\vec{q},\lambda} $ of photons of polarization
   $\lambda$ and momentum ${\vec q}$ one  defines the quantized transverse e.m. vector potential 
 
\begin{equation}
\vec{A}_{\bot}(\vec{x})=\sum_{\lambda=1,2}\sqrt{\frac{\hbar c}{\Omega}}\sum_{\vec{q}}\frac{1}{\sqrt{|\vec{q}|}}\vec{e}_{\vec{q}}^{(\lambda)}
e^{-\imath\vec{q}\vec{x}}\left(b_{\vec{q},\lambda}+ b_{-\vec{q},\lambda}^{+}\right)
\label{Aquant}
\end{equation}
 taken with periodical boundary conditions. This definition brings the photon part of the Hamiltonian to a diagonal form. Here the bosonic commutators are 
\[
\left[b_{\vec{q},\lambda},b_{\vec{q}',\lambda'}^{+}\right]=\delta_{\vec{q},\vec{q}}'\delta_{\lambda\lambda'}
\]
and the unit vectors $\vec{e}_{\vec{q}}^{(\lambda)}$ are orthogonal  to the wave vector  $\vec {q}$ and to each other  
\[
\vec{q}\vec{e}_{\vec{q}}^{(\lambda)}=0;\qquad \vec{e}_{\vec{q}}^{(\lambda)}\vec{e}_{\vec{q}}^{(\lambda')}=\delta_{\lambda\lambda'};
\qquad\vec{e}_{\vec{q}}^{(\lambda)}=\vec{e}_{-\vec{q}}^{(\lambda)};\qquad(\lambda,\lambda'=1,2) \enspace .
\] 
With these ingredients and  normal ordering the operators  one gets the non-relativistic QED  Hamiltonian 
 \begin{eqnarray}
 H^{QED} & = & \sum_{\vec{q},\lambda}\hbar\omega_{q}b_{\vec{q},\lambda}^{+}b_{\vec{q},\lambda} \label{HQED}
 \\
 &+&\left. N\left[\frac{1}{2m}\left(-\frac{\hbar}{\imath}\nabla \psi^+({\vec x}) +\frac{e}{c}{\vec A}_\bot({\vec x})\psi^+({\vec x})\right)
\left(\frac{\hbar}{\imath}\nabla\psi +\frac{e}{c}{\vec A}_\bot({\vec x})\psi\right)\right] 
+\frac{1}{2}eV\psi^*\psi \right]  \nonumber
\\ 
 & + & \frac{1}{2}\int d\vec{x}\int 
 d{\vec x}'\psi^{+}(\vec{x})\psi^{+}(\vec{x}')\frac{e^{2}}{|\vec{x}-\vec{x}'|}\psi(\vec{x}')\psi(\vec{x})\enspace ,\nonumber
\end{eqnarray}
 where, according to the general recipe of second quantization  a normal ordering  $N(... )$  had to be introduced also with respect to the photon creation and annihilation operators $b_{\vec{q},\lambda}^{+} $,  $b_{\vec{q},\lambda}$ and the photon frequency is $\omega_q= c|q|$. 
 
This nonrelativistic QED Hamiltonian coincides with the standard  one obtained directly  from the second quantized Hamilton operator of electrons interacting with a classical electromagnetic field in the Coulomb gauge, after the quantization of the transverse vector potential  according to  Eq. \ref{Aquant} and adding the energy of the photons, as it is given for example in \cite{Holstein}. 

We have shown, how a field theoretical treatment allows the use of the Lagrange formalism in deriving the non-relativistic quantum mechanical many-body theory of charged particles interacting with photons  avoiding the problems linked to point-like classical charges.

\end{document}